\documentclass[prb,aps,twocolumn,amsdtx,showpacs]{revtex4}
\usepackage{graphicx}
\usepackage{amsmath}
\usepackage{amssymb}

\newcommand{\bleq}{\ifpreprintsty
                   \else
                   \end{multicols}\vspace*{-3.5ex}{\tiny
                  \noindent\begin{tabular}[t]{c|}
                  \parbox{0.493\hsize}{~} \\ \hline \end{tabular}}
                   \fi}
\newcommand{\eleq}{\ifpreprintsty
                 \else
                   {\tiny\hspace*{\fill}\begin{tabular}[t]{|c}\hline
                    \parbox{0.49\hsize}{~} \\
                    \end{tabular}}\vspace*{-2.5ex}\begin{multicols}{2}
                    \fi}
\newcommand{\bcols}{\ifpreprintsty\else\begin{multicols}{2}\fi}
\newcommand{\ecols}{\ifpreprintsty\else\end{multicols}\fi}

\newcommand \beq  {\begin{equation}}
\newcommand \eeq  {\end{equation}}
\newcommand \bea {\begin{eqnarray} }
\newcommand \eea {\end{eqnarray}}

\begin{document}
\title{Origin and consequence of an unpinned helical magnet: application to partial order in MnSi under pressure}

\author{John M. Hopkinson}
\affiliation{Brandon University, Brandon, Manitoba, Canada, R7A 6A9}
\author{Hae-Young Kee}
\affiliation{60 St.George St., University of Toronto, Toronto, Ontario, Canada}
\affiliation{Korea Institute for Advanced Study, Seoul 130-722, Korea}
\date{\today}
\begin{abstract}
We study a classical ferromagnetic Heisenberg model in the presence of Dzyaloshinskii-Moriya interactions on the corner-shared triangle lattice formed by the Mn sites of MnSi.  We show that a sizable spin helicity can be obtained only when  the microscopic Moriya vectors lie parallel to the Mn-Mn bonds.  Further, such vectors are shown to produce an unpinned helical order characterized by a particular ordering wavevector magnitude but unpinned direction, dubbed partial order, at physically realizable temperatures. 
A consequence of such an unpinned helical ordering is that the neutron scattering intensity is sharply peaked at this wavevector magnitude.  The surface formed by connecting these wavevectors is a sphere, around which the neutron scattering weight is spread. 
We further show that the observed neutron scattering intensity can be anisotropic along this surface and that this anisotropy is dependent on the experimentalist's choice of lattice Bragg peak through a geometric factor.  A neutron scattering measurement near the Bragg point ($\frac{2\pi}{a}$,$\frac{2\pi}{a}$,0) naturally leads to a highest intensity along the (1,1,0) direction consistent with the observed anisotropy in MnSi [Pfleiderer {\it{et al.}} Nature {\bf{427}} 227 (2004)]. A possible mechanism for pinning the helical order, and a way to distinguish an ordered and a partially ordered state in the context of neutron scattering are discussed.   

\end{abstract} 

\maketitle



{\it{Introduction:}} A recent paradigm for the discovery of new states of matter is the investigation of systems in close proximity to a quantum critical point (qcp). 
It was thus of great interest when high pressure neutron scattering{\cite{pflei}} of one of the most studied binary systems, the helimagnet MnSi, showed evidence of a new type of magnetic state arising near a closely avoided qcp.  Specifically, the periodicity of the spin helices was found to be fixed while their relative orientations had depinned, leading to a very disperse structure factor transverse to a well-defined radial ordering wavevector magnitude dubbed ``partial order''.  It has not been understood how such an anisotropic neutron scattering intensity arises from a partially ordered state.  Starting from the same pressures, rather than exhibiting a quick recovery of $\Delta\rho \sim T^{2}$ as expected from spin fluctuation theory close to an itinerant qcp, resistivity measurements were found to exhibit a non-Fermi liquid character $\Delta\rho \sim T^{\frac{3}{2}}$ over an extended pressure range{\cite{nicholas1,Jaccard}.  Moreover, the application of a small magnetic field, as might be expected to pin the helical axis, was found{\cite{nicholas1,nicholasII}} to lead to a low temperature recovery of Fermi liquid character, leading one to suspect an intimate relation between the proposed partially ordered state and the unusual non-Fermi liquid resistivity prevalent in zero field.

 It has long been believed that the essential physics present in itinerant helimagnets on this lattice, including MnSi at ambient pressures, is the result of the interplay between itinerant ferromagnetic order and Dzyaloshinskii-Moriya (D-M) spin-orbit coupling terms. These tend to favor canted spin structures and are present because of the lack of inversion symmetry of the lattice{\cite{naganishi, bak}}.  A Landau-Ginzburg(LG) based phenomenological approach has led 
to many successes including the prediction of magnetic field induced wavevector reorientations{\cite{walker}}. 
 The LG free energy is expanded in terms of slowly-varying spin densities (${\bf S}({\bf{r}})$ in terms of which the DM
interaction  is given by ${\bf S}({\bf r}) \cdot \left( \nabla  \times {\bf S}({\bf r})\right)$.
In such a continuum system,  DM interaction  has full rotational symmetry.
As a result, the direction of ordering wave-vector is determined by a further anisotropic second
order gradient term.\cite{bak}
On the other hand, on a lattice, the DM interaction is given by ${\bf D}_{i j} ({\bf S}_i \times
{\bf S}_j)$, where ${\bf D}_{ij}$ called the Moriya vector is defined on the link
between two sites $i$ and $j$.
In general, the DM interaction breaks spin rotational symmetry due to a particular direction
of Moriya vector, resulting in a preferential direction of the spin wave ordering wave-vector.  The successful application of LG theory in the description of the physics of MnSi up to now lessened the motivation to study of the possible microscopic Moriya vectors of this lattice structure.
However, an understanding of the phase transition from an ordered helical state to
the proposed partially ordered state as a function of applied pressure in this strongly correlated material
requires a study beyond the standard LG theory\cite{bak}.

More recently, to describe the partially ordered state, the addition of a pseudoscalar order parameter was argued{\cite{Tewari}} to lead to a ``blue quantum fog'' with properties of a chiral liquid.  On the other hand, attempts to explain an observed shift of weight from the (111) direction of the ambient pressure ordered helimagnet, to the apparent favoring of the (110) direction at high pressures, has led to theories which favor a weakly ordered state.  This is argued to result from interactions between helimagnetic structures leading either to a new type of spin crystal{\cite{Binz,shah}} or a lattice of skyrmions{\cite{bogdanov}}.  

 In this paper, we attempt to provide a microscopic model for helical ordering on the corner-shared triangle lattice formed by the Mn sites of MnSi, and to investigate one consequence of partial order in the context of the observed anisotropic neutron scattering intensity.   We find that for reasonable spin-orbit couplings, the strongest component of the Moriya vectors must lie parallel to the Mn-Mn bonds, and favors nearly ideal simple helimagnetic structures with all spins perpendicular to the axis of propagation.  However, in contrast to current folklore, at accessible temperatures the D-M interaction does not pin a direction of the helical ordering wavevector, 
 but determines only its magnitude.  A possible mechanism of pinning these helices to lie along the (1,1,1) direction at ambient pressure will be discussed. 
At high pressures, while the observation of high neutron scattering intensity along the (110) direction has stimulated ideas of crystal lattices featuring exotically ordered states, we show that such an anisotropic intensity can originate from a simple geometrical form factor, given a partially ordered state of unpinned helices.  To confirm such an Occam's razor partial order further neutron scattering measurements relative to lattice Bragg peaks other than ($\frac{2\pi}{a}$,$\frac{2\pi}{a}$,0) are suggested\cite{note}.

{\it{Moriya vectors:}} The D-M interaction first came to wide attention to explain weak FM moments in largely antiFM $\alpha$-Fe$_2$O$_3$ crystals by Dzyaloshinskii{\cite{Dz}}, but had been introduced several years earlier by Stevens{\cite{stevens}} where it was derived as a consequence of the inclusion of spin-orbit coupling in the Heisenberg model.  Moriya{\cite{moriya}} included spin-orbit coupling in a consideration of the superexchange mechanism to provide the first microscopic derivation,
introducing $\overrightarrow{D}_{ij}$ (the Moriya vector) which is proportional to the first power of the spin-orbit coupling and an antisymmetric vector, and $\overleftrightarrow{\Gamma}_{ij}=\frac{\overrightarrow{D}_{ij}\otimes\overrightarrow{D}_{ij}}{2J_{ij}}$  a symmetric tensor of second order in the spin-orbit coupling. 
For collinear spins, Yildirim et al{\cite{Yildirim}} showed that the symmetric term can be of the same order of magnitude as the antisymmetric term, correcting Moriya's expression for the single U Hubbard model to read,
\begin{eqnarray}
H =\sum_{\langle ij\rangle}&&((J_{ij}-\frac{|\overrightarrow{D}_{ij}|^2}{4J_{ij}})\overrightarrow{S}_i\cdot\overrightarrow{S}_j+\overrightarrow{D}_{ij}\cdot(\overrightarrow{S}_i\times\overrightarrow{S}_j)\nonumber\\&+&\overrightarrow{S}_i\cdot \frac{\overrightarrow{D}_{ij}\otimes\overrightarrow{D}_{ij}}{2J_{ij}}\cdot\overrightarrow{S}_j)\label{equation2}
\end{eqnarray}
where $[A\otimes B]_{\mu \nu} = A_{\mu}B_{\nu}$ and $J_{ij}$ is the regular Heisenberg coupling.  As the magnetic interactions in MnSi are thought to be predominantly ferromagnetic in origin, below we keep both terms linear and quadratic in $D_{ij}$, however our main results do not appreciably change if we drop the contributions of the symmetric spin couplings.
\begin{figure}
\includegraphics[scale=0.5]{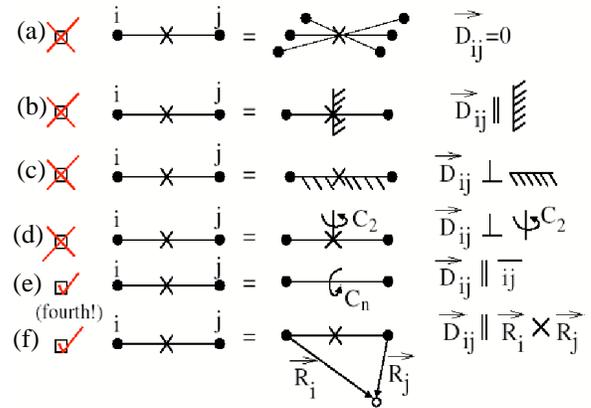}
\caption{\label{clstfig}(Color online)Moriya's rules (a)-(e) for the determination of the microscopic Moriya vectors\cite{moriya} in insulating materials due to superexchange interactions and (f) due to magnetic impurities at sites $i$ and $j$ in itinerant systems\cite{fertlevy,levyfert}.  Such vectors live on the bonds at the midpoint between atoms at sites $i$ and $j$.  Directions are assigned as shown when the crystal symmetries lead to: (a) the midpoint of $R_{ij}$ being a centre of inversion, (b) the midpoint of $R_{ij}$ being a mirror plane perpendicular to $R_{ij}$, (c) a mirror plane parallel to $R_{ij}$, (d) a 2-fold rotation axis perpendicular to $R_{ij}$ through the midpoint of $R_{ij}$, (e) an n-fold axis parallel to $R_{ij}$, and (f) an itinerant atom with a spin-orbit coupling living a distance $R_i$ and $R_j$ respectively from local moments at sites $i$ and $j$. Symmetries $(a)-(d)$ are not found for the MnSi lattice, which features a $P2_13$ symmetry.  Case (e) applies, but only for $ij$ being fourth nearest neighbours.  \label{figure10}}
\end{figure}


To understand how to choose Moriya vector directions, it is useful to recall that
MnSi is  an itinerant magnetic system which shows features reminiscent of strongly correlated systems.
In particular, one finds an enhanced effective mass $~ 10 m_e$\cite{taillefer}, a large effective moment at high temperatures
(~2.2 $\mu_B$/Mn\cite{wernick}, consistent with spin-$\frac{5}{2}$ local moments at each site, above $T_c$),
and a reasonable ordered magnetic moment (~0.4 $\mu_B$/Mn)\cite{carbone,thessieu}.
It has been argued that x-ray absorption measurements exhibits strong correlation effects.\cite{carbone}
Therefore, it is plausible that at low temperatures the electrons at each site have a dual character, producing both a small local moment and itinerant conduction electrons.  In this case, magnetic RKKY interactions may arise between the local moments on each Mn site
through these itinerant conduction electrons.  The magnetic properties of such a model could find description in the language of the extended
Heisenberg model.
As the dominant magnetic interaction appears to be ferromagnetic in origin,
as evidenced by the positive Curie offset of the magnetic susceptibility,  we here consider the further simplification of keeping only the order 1 terms of such an expansion, 
 the nearest neighbour FM and DM interactions.
One possible mechanism of helical ordering with wave-vector along $(110)$ will be discussed below.

\begin{figure}
\includegraphics[scale=0.34]{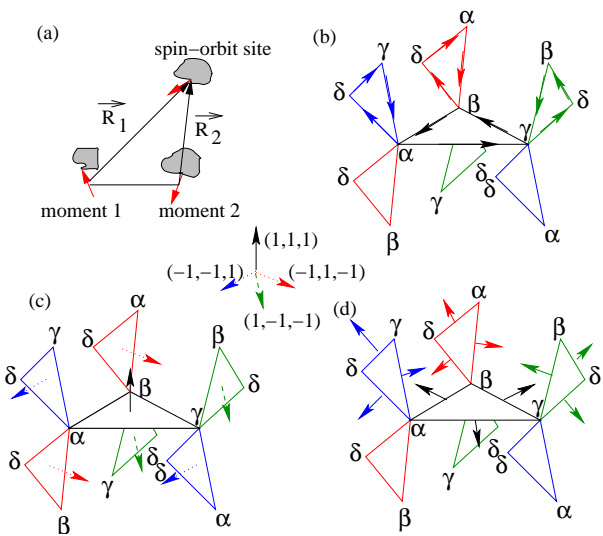}
\caption{\label{clstfig}(Color online) a) The RKKY interaction of Fert and Levy chooses Moriya vectors parallel to the cross product between the direction vectors from the local moment site to the spin-orbit site; b)-d) An orthogonal basis set of Moriya vector choices consistent with the lattice symmetries: b) parallel to the bonds; c) parallel to the normal vectors, consistent with nearest neighbor RKKY contributions ``RKKY I''; d) the cross-product between b) and c), consistent with further neighbor RKKY contributions ``RKKY II''.\label{figure1}}
\end{figure}

In itinerant magnetic systems, it was shown that the RKKY interaction 
is modified by the spin-orbit coupling\cite{fertlevy,levyfert}, generating microscopic Moriya vectors between local moment sites.  Their analysis was applied to non-magnetically doped CuMn$_x$ spin glass alloys in the dilute impurity limit.  They found, 
that $\overrightarrow{D}_{ij}\propto\frac{\lambda(\hat{R}_i\times\hat{R}_j)}{R_iR_jR_{ij}}$,
where 
$\lambda$ is the spin-orbit coupling constant at a site of distance $R_i$ and $R_j$ respectively from two magnetic sites separated by distance $R_{ij}$ as shown in Fig. \ref{figure10} (f).
For lattices like MnSi, where Moriya's rules as shown in Fig. \ref{figure10} do not determine the direction of any nearest neighbor Moriya vectors, this allows one to construct a basis of directions consistent with the lattice symmetries, and potentially provides a mechanism for the realization of such an interaction.  
It is interesting to note that on the corner-shared tetrahedral lattice common to pyrochlore systems the nearest neighbor RKKY interaction determination agrees with the direction predicted by Moriya's rules{\cite{elhajal}}. 

Let us consider all possible sets of Moriya vectors consistent with the lattice symmetries. The lattice structure of the magnetic Mn sites is simple cubic with a four site basis, having an overall $P2_13(T^4)$ symmetry. Sites can be labelled:  $\{\alpha,\beta,\gamma,\delta\}=\{(u+\frac{1}{2},\frac{1}{2}-u,1-u),(1-u,u+\frac{1}{2},\frac{1}{2}-u),(\frac{1}{2}-u,1-u,u+\frac{1}{2}),(u,u,u)\}$ with $u_{Mn}=0.138$.  The silicon sites form a similar corner-shared triangle lattice structure with $u_{Si}=0.854$. Fig. \ref{figure1} (a) is a cartoon of the rule of Fig. \ref{figure10} (f) as it would be applied to bonds between local moments on neighboring Mn sites mediated by itinerant conduction electrons experiencing the spin-orbit scattering by a third Mn site.  Basis vectors of Fig. \ref{figure1}(c) arise from considering the spin-orbit RKKY contribution from the third Mn site on each triangle, components along Fig. \ref{figure1}(d) would arise from considering the closest Si atom as the spin-orbit site.  An important limitation of our analysis is that the form of the Moriya vectors of Ref. \onlinecite{fertlevy,levyfert} was determined in the dilute impurity limit, whereas its application to MnSi would be in the dense impurity limit.  As such, Fig. \ref{figure1}(b) completes an orthogonal basis with Moriya vectors possibly of non-RKKY origin.  While Moriya vectors as shown in Fig. \ref{figure1} (b) are consistent with all Moriya rules and the lattice symmetries of MnSi, the microscopic derivation of such a possibility remains an interesting open problem.

{\it{Partial order:}} For nearest neighbor ferromagnetic couplings ($J_{ij} < 0$) between classical spins we have studied the low energy spectrum of the Hamiltonian given in Eq. \ref{equation2}.  We have used both mean field\cite{reimers,hopkinson} and real space rigid rotor minimization\cite{hopkinson} methods.  The former method tends to over-estimate the likelihood of a spin structure to cant, but has no difficulty exploring extremely long period structures, and has been used to gain a qualitative feel for the relative strength of different direction couplings.  In the latter method, we impose periodic boundary conditions on finite size structures with a given periodicity which allows us to access reasonably small ordering wavevectors\cite{hopkinson}.  

\begin{figure}
\includegraphics[scale=0.23]{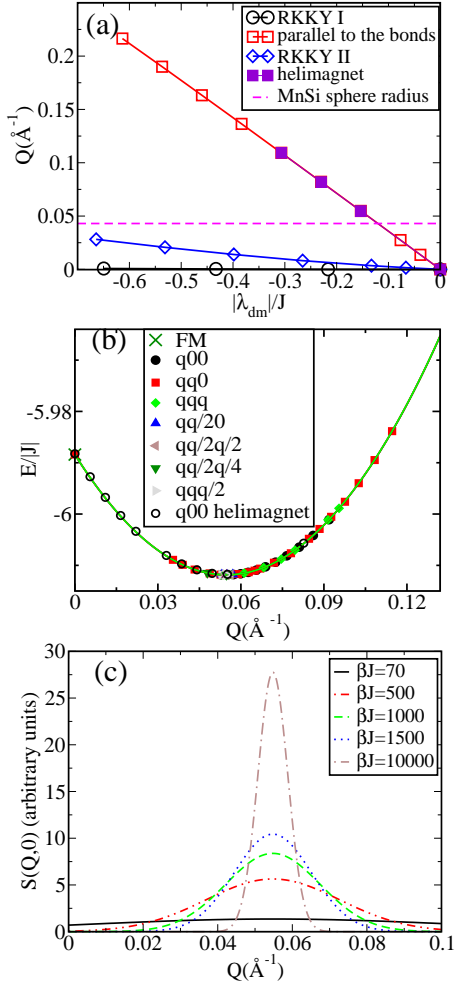}
\caption{\label{clstfig}(Color online) (a) Ordering wavevector magnitude as a function of the D-M interaction for choices of the Moriya vectors shown in Fig. \ref{figure1}, where $|\lambda_{dm}|$ is set by the strength of $|D_{ij}|$.  Moriya vectors taken parallel to the bonds lead to simple spirals which can account for a period as short as MnSi; those of RKKY I do not cant spins; while those of RKKY II lead to internal structure and very weak canting. (b) For $|\frac{\lambda_{dm}}{J}|=0.25\sqrt{(2u_{Mn}-\frac{1}{2})^2+(2u_{Mn})^2+(\frac{1}{2})^2}\approx 0.153$, $Q\equiv\frac{1}{a}\sqrt{q_x^2+q_y^2+q_z^2}$ and Moriya vectors parallel to the bonds, the dispersion is to a very good approximation given by a degenerate sphere in momentum space.  Filled symbols: rigid rotor solutions, solid lines along q00, qq0, qqq: mean field, and open circles: pure helimagnet.  (c) At low temperatures radial cuts of the structure factor can become sharply peaked along any direction in comparison to transverse cuts for energetic reasons as discussed in the text.\label{figure2}}
\end{figure}

 In Fig. \ref{figure2} (a), we show mean field results for the magnitude of the ordering wavevector for the three basis vectors.  It is readily apparent that, given the size of the ordering wavevector of MnSi, the dominant contribution must arise from a choice of Moriya vectors parallel to the bond.  This is confirmed by our rotor model calculations which find no evidence of canting when the Moriya vectors are chosen to lie in the primary RKKY I direction of Fig. \ref{figure1} (c), and internal structure and only very weak canting when they are chosen in the secondary RKKY II direction of Fig. \ref{figure1} (d).  In contrast, a choice of Moriya vectors parallel to the bonds, as shown in Fig. \ref{figure1} (b), leads to almost perfect helimagnets with spin structures pinned in the plane orthogonal to the propagation direction.  

Using a set of Moriya vectors lying parallel to the bonds, we investigate the directional energy dependence for helimagnetic wavevectors, $\overrightarrow{Q}$.  As shown in Fig. \ref{figure2} (b) the spectrum along an infinite number of directions, $\hat{Q}$, is found to be degenerate within a precision of ${\mathcal{O}}(3\times 10^{-6}J)$\cite{noteen}.  It is important to note that the wavevectors delineating members of this manifold lie on the surface of a sphere of radius $|\overrightarrow{Q}|$ in momentum space.    A pure helimagnet as,
\begin{equation}
s(x)=\sin(\overrightarrow{Q}\cdot\overrightarrow{r})\hat{a}+\cos(\overrightarrow{Q}\cdot\overrightarrow{r})\hat{b},\label{equation3}
\end{equation}
 with $\frac{\hat{a}\times\hat{b}}{|\hat{a}\times\hat{b}|}=\hat{Q}$,  where $\hat{a}$ and $\hat{b}$ are orthogonal unit vectors is a reasonable approximation to the ground state spin structure as shown with open circles, lying only ${\mathcal{O}}(8\times 10^{-5}J)$ higher in energy.  Mean field calculations (which feature a soft spin constraint) are found to provide a reasonable approximation to the energetics of the true minimized spin structures at small Q as shown by the solid lines of Fig. \ref{figure2} (b), allowing a smooth interpolation of the low energy states available to a finite temperature system.  This means that we can approximate the probability of a given state as $P(\epsilon_k)=e^{-\beta\epsilon_k}/(4\pi\sum_k(k^2\Delta ke^{-\beta\epsilon_k}))$.  The square of this probability, as seen in Fig. \ref{figure2} (c), is the dominant contributor to the radial structure factor.

{\it{Anisotropic structure factor:}} 
An interesting question is whether the anisotropic neutron scattering intensity strongly favoring the (110) direction under pressure in MnSi can be understood to arise from a partially ordered state, that is, from an unpinned helical order.  Here we show that an anisotropic intensity can be understood to originate from a geometric factor present due to the dipolar nature of the neutron-electron interaction.  The interaction between the magnetic moments of the neutrons and electron spin leads to the form $H_{int}(s_j)\propto e^{-i\overrightarrow{\kappa}\cdot\overrightarrow{R}_j}\overrightarrow{\mu}_n\cdot(\hat{\mathcal{Q}}\times\overrightarrow{s}_j\times\hat{\mathcal{Q}})$ where $\overrightarrow{\mu}_n$ is the neutron magnetic moment, $\hbar \overrightarrow{\mathcal{Q}}$ the momentum transfer, and $\overrightarrow{s}_j$ the spin at site j.  The elastic contribution to the differential cross-section is found in the Born approximation to be{\cite{Jensen}},
\begin{equation}
\frac{d\sigma}{d\Omega}\propto\sum_{\alpha,\beta}(\delta_{\alpha\beta}-\hat{\mathcal{Q}}_{\alpha}\hat{\mathcal{Q}}_{\beta})\sum_{ij}\langle s_{i\alpha}\rangle\langle s_{j\beta}\rangle e^{-i\overrightarrow{\mathcal{Q}}\cdot(\overrightarrow{R}_i-\overrightarrow{R}_j)}\label{equation4}
\end{equation}
where $\alpha,\beta$ run through $\{x,y,z\}$.   In Eq. \ref{equation3}, we found that a reasonable approximation to the ground state was a pure spin spiral with a small wavevector, $Q$. 
As the measured MnSi data{\cite{pflei}} lie about the $(110)$ Bragg peak we have ${\overrightarrow{\mathcal{Q}}}=\overrightarrow{Q}+(\frac{2\pi}{a},\frac{2\pi}{a},0)$, and one expects that both an atomic form factor and an angular dependence should arise.   Assuming the form of Eq. \ref{equation3} at zero temperature, we find the structure factor to be proportional to,
\begin{equation}
(1-\cos(8\pi u))(1-\frac{1}{2}((\hat{a}\cdot\hat{{\mathcal{Q}}})^2+(\hat{b}\cdot\hat{\mathcal{Q}})^2).\label{equation5}
\end{equation} 
Note that the geometric form factor in Eq. \ref{equation4} indicates that unpolarized neutron scattering is insensitive to the component of a spin structure parallel to the momentum transfer of the scattering neutrons.  Hence, for a truly degenerate simple helimagnet measured with respect to the (0,0,0) lattice Bragg peak  one expects to see no angular structure, while measurements with respect to the $(1,1,0)$ lattice Bragg peak will generate diffuse intensities over the surface of a sphere in momentum space which peaks along $(1,1,0)$.   This is illustrated in Fig. \ref{figure3}. 

\begin{figure}
\includegraphics[scale=0.3]{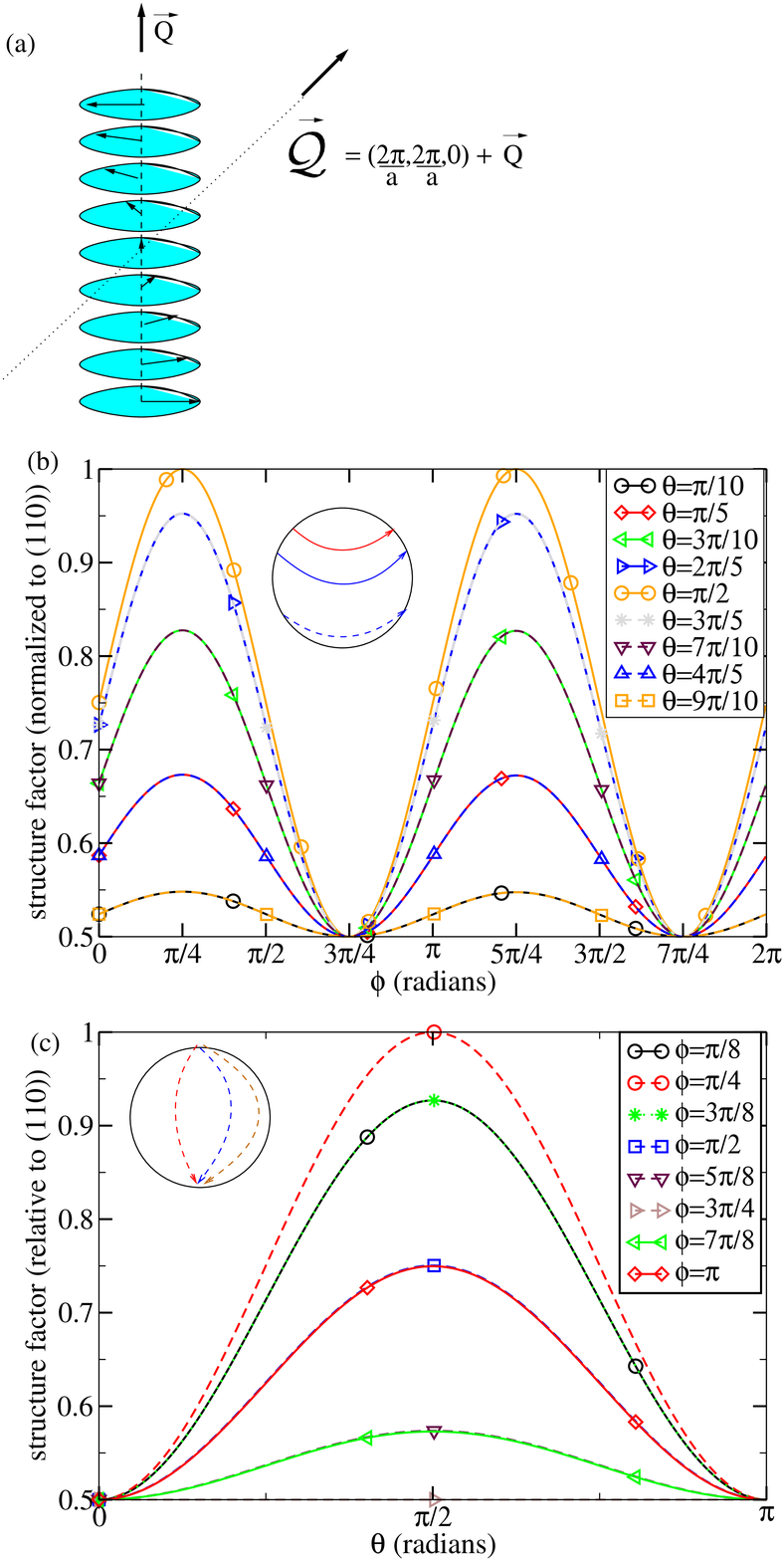}
\caption{\label{clstfig}(Color online) (a) While the spin plane (aqua) of the helimagnet pins perpendicular to the ordering wavevector ${Q}$, 
this spin plane may not be perpendicular to the momentum transfer of the neutrons,  ${\mathcal{{\vec{Q}}}}={\vec{Q}}+(\frac{2\pi}{a},\frac{2\pi}{a},0)$.  Note that constructive interference leads to a magnetic signal at all points related to ${\vec{Q}}$ by a reciprocal lattice vector. (c)-(d) Variation of the structure factor arising due to the geometric factor (Eq. \ref{equation5}) about the lattice Bragg peak ($\frac{2\pi}{a}$,$\frac{2\pi}{a}$,0) assuming a pure spiral ground state of the form Eq. \ref{equation3}, with $|\vec{Q}|$=0.043\AA. $\theta$ and $\phi$ are chosen such that $\vec{Q}=|\vec{Q}|$($\sin(\theta)\cos(\phi)$,$\sin(\theta)\sin(\phi)$,$\cos(\theta)$).  Note the maximum occurs at (1,1,0) and (-1,-1,0) but not at (1,0,1), (0,1,1).\label{figure3} }
\end{figure}

{\it{Discussion:}} The nature of the high pressure state of MnSi remains a topic of active experimental debate.  While some have argued that MnSi undergoes a first-order phase transition at quite low pressures{\cite{petrova}, the predominant focus has been on understanding the detailed magnetic properties of the observed signature of partial order.  While nuclear magnetic resonance studies of powdered samples found decreasing magnetic moments well above the critical pressure above which the helices have depinned{\cite{yu}}, a careful muon spin relaxation study{\cite{uemura}} of single crystals did not find ordered moments.  Rather, Uemura {\it{et al.}}{\cite{uemura}} argued for a ``slow but dynamic order'' as responsible for the diffuse neutron scattering signal observed, suggesting that the spins must fluctuate on a timescale between $10^{-11}$ and $10^{-10}$s.   Most interestingly, a sizable contraction of the lattice has been observed by neutron Larmor diffraction{\cite{larmor}} to be concomitant with the onset of non-Fermi liquid behaviour at high pressures, ruling out most quantum critical scenarios (ex. magnetic rotons{\cite{schmalian}}).  It would appear that this lattice contraction is responsible for the depinning of the helimagnetic order.

In this light, we briefly comment on the pinning of the ambient pressure (1,1,1) ordered state.  The local environment of the Mn site is surrounded by 7 Si atoms, with the shortest Mn-Si distance along the local (1,1,1) axis.  It therefore seems plausible that crystal/ligand field effects might produce a local (1,1,1) anisotropy of the spin structure.  One term which might arise from a full consideration of spin-orbit interactions in this context is of the form, $-\zeta \sum_{i,\nu}(\overrightarrow{s}_{i,Ising}^{\nu})^4$ where the sum runs over the number of unit cells and the four sites within the unit cell, $\overrightarrow{s}_{i,Ising}^{\nu}=(\overrightarrow{s}_i\cdot\overrightarrow e_{\nu})\overrightarrow e_{\nu}$ and $\overrightarrow e_{\nu}$ is the unit vector associated with the local (1,1,1) anisotropy at site $\nu$.  Assuming this term is small, we can evaluate the energy change such a term would introduce to the simple helimagnet.  For $\zeta>0$ this term is found to lower the energy of the 100, 110 and 111 directions by -6$\zeta$, -6.5$\zeta$ and -8$\zeta$ respectively per unit cell, effectively favoring (1,1,1).  The increase of pressure would likely increase the overlap of the various Si atoms with the Mn sites, reducing the strength of pinning.

We have shown that the geometric form factor provides one possible explanation of the higher intensity along the (1,1,0) direction measured by neutron scattering{\cite{pflei}} about the ($\frac{2\pi}{a},\frac{2\pi}{a},0$) Bragg peak.  However, we cannot rule out the possibility that additional terms may be present and weakly favor ordered helices along (1,1,0).  In particular, we have shown\cite{us} that the addition of antiFM 2nd and 3rd nearest neighbors generically leads to a (1,1,0) ordered helical state.  We expect that this (1,1,0) ordered helical state will be robust in the presence of the DM interaction, as the DM interaction does not pin the direction.\cite{afterus}  To distinguish these two different proposals, we propose further neutron scattering around different Bragg points such as ($\frac{2\pi}{a},0,0$) is desirable--in the partially ordered state a higher intensity moves to (1,0,0) while in a (1,1,0) ordered helical state, the neutron scattering intensity remains peaked at (1,1,0).  Even within the existing study, it is our belief that the ``rotary electric shaver'' picture of the intensity created by mathematica for publication by Pfleiderer {\it{et al.}}\cite{pflei}, while consistent with their data, may have inadvertantly symmetrized their intensities, as we would not expect a similar enhancement near (0,1,1) and (1,0,1) arising solely from the partial order here considered. 


{\it{Summary:}} To conclude, we have shown that the dominant contribution of the Moriya vectors in MnSi must lie parallel to the Mn-Mn bonds.  This leads to a dispersion minimum at a particular helix pitch independent of the direction of order.  The corresponding momentum space sphere has a much smaller energetic spread than accessible temperatures, leading to a sharply peaking radial structure factor, and a very disperse transverse structure factor.  We offer a natural explanation for the higher intensity of the structure factor at (1,1,0)  as observed experimentally.  The application of weak magnetic fields would be expected to break this degeneracy, pin the orientation of the helimagnets, and it is tempting to argue that the pinning of the helical order direction leads to 
the recovery of $\Delta\rho\sim T^2$ behavior seen.  


{\it{Acknowledgments:}} J.H. thanks S. Isakov, K. Sengupta and B. Binz for useful discussions.  We thank C. Pfleiderer for additional information regarding the neutron scattering analysis carried out in Ref. \onlinecite{pflei}. This work was supported by an NSERC PDF, an NSERC CRC, NSERC and the CIAR.  

[Note Added]
After a completion of our work, the author of Ref. \onlinecite{pflei} agreed that
that magnetic scattering near a nuclear Bragg spot is subject to a selection rule, which
has not been discussed in Ref. \onlinecite{pflei}.
However, he  pointed out that Fig. 2 (b) and (d) in Ref. \onlinecite{pflei}
show that the intensity drops off much faster than what one would expect from the effect of
geometrical form factor in unpinned helicial state.
While we show that the intensity drops to 50\%, when the azimutal angle, $\phi$ becomes $ \pm \pi/2$
at a fixed $\theta = \pi/2$ as shown by a line with circles in Fig. \ref{figure3} (b), the  experimentally observed intensity
drops to almost 0  near $\pm \pi/6$, indicating neither an unpinned helix nor an ordered state.  Other
(110) lattice points exhibit the maxium intensity which have not been reported in Ref. \onlinecite{pflei}.
These confirm that further studies such as neutron scattering around
other Bragg lattice points are needed to determine the detailed nature of the partially ordered state.




\begin{thebibliography}{00}
\bibitem{note} Here $a=4.56$\AA is the lattice constant for MnSi.
\bibitem{noteen}  Within mean field theory the energetic spread around the 
surface of this sphere is seen to be tiny ${\mathcal{O}}(9.4\times 10^{-7}J)$.
The largest deviation from mean field theory for our minimized spin structures
(rotors) is $ {\mathcal{O}}(5\times 10^{-6}J)$, which represents an upper 
estimate of our precision.  The energetic spread among minimized spin 
structures is seen to be $ {\mathcal{O}}(3.5\times 10^{-6}J)$
For the case $|\lambda_{dm}/J| = 0.25\sqrt{(2u_{Mn}-\frac{1}{2})^2+(2u)^2+(\frac{1}{2})^2}$.
\end{thebibliography}
\end{document}